# What is Implementation Science; and Why It Matters for Bridging the Artificial Intelligence Innovation-to-Application Gap in Medical Imaging

Ahmad Fayaz-Bakhsh, MD, PhD[1,2,*], Janice Tania, B.S.[3], Syaheerah Lebai Lutfi, PhD[4,5], Abhinav K. Jha, PhD[3,6], Arman Rahmim, PhD, DABSNM[7,8]

[1] School of Public Health, Tehran University of Medical Sciences, Tehran, Iran
[2] Self-Care Academic Research Unit (SCARU), Department of Primary Care & Public Health, School of Public Health, Imperial College, London, UK
[3] Department of Biomedical Engineering, Washington University in St. Louis, St. Louis, MO, USA
[4] Medical Education and Informatics, Sultan Qaboos University, Muscat, Sultanate of Oman
[5] School of Computer Sciences, Universiti Sains Malaysia, Penang, Malaysia
[6] Mallinckrodt Institute of Radiology, Washington University in St. Louis, St. Louis, MO, USA
[7] Departments of Radiology and Physics, University of British Columbia, Vancouver, Canada
[8] Department of Basic and Translational Research BC Cancer Research Institute, Vancouver, Canada

*Corresponding Author: Ahmad Fayaz-Bakhsh, MD, PhD, School of Public Health, Tehran University of Medical Sciences; Email: fayaz@tums.ac.ir

**Key Points**

- The transformative potential of artificial intelligence (AI) in medical Imaging (MI) is well recognized. Yet despite promising reports in research settings, many AI tools fail to achieve clinical adoption in practice. In fact, more generally, there is a documented 17-year average delay between evidence generation and implementation of a technology[1].
- Implementation science (IS) may provide a practical, evidence-based framework to bridge the gap between AI development and real-world clinical imaging, to shorten this lag through systematic frameworks, strategies, and hybrid research designs.
- We outline challenges specific to AI adoption in MI workflows, including infrastructural, educational, and cultural barriers.
- We highlight the complementary roles of effectiveness research and implementation research, emphasizing hybrid study designs and the role of integrated KT (iKT), stakeholder engagement, and equity-focused co-creation in designing sustainable and generalizable solutions.
- We discuss integration of Human-Computer Interaction (HCI) frameworks in MI towards usable AI.
- Adopting IS is not only a methodological advancement; it is a strategic imperative for accelerating translation of innovation into improved patient outcomes.

## Introduction

Artificial intelligence (AI)-based methods and solutions continue to garner considerable attention in the field of MI. Across the MI landscape, promising capabilities are demonstrated in workflow optimization, data acquisition, image reconstruction, image enhancement, lesion detection and segmentation and computer aided diagnosis. These innovations have shown potential in various applications such as neuroimaging,[2] cardiacimaging,[3,4] oncologicalimaging,[5–8] to name a few. Such innovations have created AI-enhanced clinical decision support systems (CDSSs)[9]. Before the recent surge in AI, however, CDSSs were being used in one form or another in the context of telemedicine[10,11], mobile health (m-Health)[12,13], electronic health record (EHR)[14,15] and office automation systems[16],and were being subjected to value assessments for many years[17].Despite digital technology advances in the years before and after AI, many validated solutions have remained stalled at the edge of clinical practice. The overwhelming majority of AI-based imaging studies remains limited to proof-of-concept investigations or retrospective validations and rarely progress into real-world deployment[2].

There is a clear gap between creating innovative solutions – especially AI in MI - and actually using them in real clinical settings. This gap shows the need for a field that does not just focus on developing new tools, but also on how to successfully adopt, integrate, and sustain them in the real world of healthcare. IS provides this missing link (to be defined in details below). It offers structured approaches to identify and address the many barriers that prevent clinical use of new technologies. In this manuscript, we explain the background and reasons behind the rise of IS, clarify key terms, and explore useful models and frameworks. We also discuss how IS connects with knowledge translation and AI evaluation methods, how it can be combined with human-computer interaction (HCI) principles, and how hybrid research designs can assess both clinical effectiveness and implementation success. We then highlight common barriers to implementing AI and offer strategies and real-world applications to overcome them. Finally, we stress the importance of collaboration, especially co-creating solutions with those who will use them.

## 1. What is Implementation Science?

Imagine you've developed a life-saving serum that is potent, effective, and backed by rigorous evidence. But unless that serum is delivered reliably to the patient, its potential is wasted. In healthcare, evidence is the serum. Implementation strategies are the delivery system. This is the core idea behind Implementation Science (IS): ensuring that validated tools, such as AI technologies in MI, actually reach patients in a usable, sustainable, and impactful way. To understand the origins and importance of IS, we must first revisit **Evidence-Based Medicine (EBM)**. EBM emerged in the early 1990s as a paradigm shift in clinical practice, emphasizing the conscientious, explicit, and judicious use of current best evidence in making decisions about the care of individual patients[18,19] As the field evolved, the concept broadened into **Evidence-Based Practice (EBP)**, extending beyond medicine to encompass more health professions. EBP emphasizes three key elements: the best available research evidence, clinical expertise, and patient values and preferences. This holistic approach acknowledges that publishing research findings alone are insufficient to improve EBP awareness[19]and to guide clinical decision-making in diverse, real-world settings. However, despite its theoretical appeal, the EBM/EBP movement has faced a significant challenge: there appears to be an enduring delay between publication growth of scientific articles, evidence generation and its integration into clinical practice. In fact, few clear relationships could be observed between local scientific article growth, economic wealth, and more uptake and implementation of the local published evidence and subsequent innovations and human development based on locally developed science.[20]It is frequently cited that it takes an average of 17 years for research evidence to become routine clinical practice.[1]Documented lags of up to a decade in areas such as telemedicine,[10,17,21] prescribing practices,[22] and interventions such as mammography screening and tobacco cessation for cancer control[23], illustrate how even well-supported evidence often fails to translate into clinical impact without deliberate system level support, effective communications, and behavioral change techniques.[22,24]

This persistent "evidence-to-practice" bottleneck reveals the limitations of EBM when applied in isolation. It assumes that once high-quality evidence is generated, clinical adoption will follow. In reality, healthcare providers and systems are influenced by a wide range of organizational, behavioral, technological, access-related and policy-related barriers.[9,25,26] These, in turn, hinder user satisfaction,[27,28] delay infrastructure improvements,[29] and obstruct provider-driven innovation and entrepreneurship,[30] which are all essential factors for the more rapid uptake of EB solutions, and in turn, more improvement in health.[31]

The field of IS provides a necessary and complementary companion to EBM/EBP to address this challenge.[32] IS is the study of methods to promote the systematic uptake of research findings and other evidence-based practices into routine practice, with the goal of improving the quality and effectiveness of health services.[33]Unlike EBP, which often assumes that individuals will adopt beneficial interventions once informed,[34] IS acknowledges that knowing something works is rarely enough, and challenges the assumption that once an intervention is proven effective, it will be adopted organically. Clinicians, administrators, and patients do not always behave in predictable, linear ways. Even proven innovations often require intentional, targeted strategies to actively push or pull systems and individuals toward meaningful change and reach their intended impact.

The field of IS draws from public health, behavioral science, organizational theory, and marketing. It employs a range of tools—theories, frameworks, and mixed-methods approaches—to identify and overcome barriers to implementation. Whether the challenge in diffusion of innovation lies in the design of the intervention, system infrastructure, provider training, or cultural context,[13,35]. IS provides structured strategies to improve adoption, fidelity, and sustainability.[33]We remind our readers to the above analogy to clarify EBP as opposed to implementation strategy.

**2. Why Does MI Need IS?**

Despite rapid advances in technology, medicine continues to face a significant gap between innovation and clinical adoption. New tools are often rigorously developed and validated, yet rarely make it into routine clinical use[36,37]. This mirrors a broader pattern in healthcare: high-quality evidence alone does not ensure timely or effective adoption in practice. In imaging, this "evidence-to-practice" gap is especially pronounced. Complex workflows, specialized infrastructure, regulatory uncertainty, and behavioral inertia among users can all hinder implementation[9,21]. Without structured strategies to overcome these barriers, even the most promising technologies may fail to benefit patients.The urgency of addressing these needs is clear: in fields such as oncology, neurology, and cardiology, where MI plays a central role in early detection and treatment planning, delays in implementation can have direct consequences on patient outcomes[38]. IS enables us to translate imaging innovation into impact—sooner, smarter, and more equitably[35].

IS offers a roadmap to close this gap. It provides imaging departments and health systems with tools to educate users, adapt workflows, and monitor real-world performance—ensuring that innovations are not only effective in controlled settings, but also usable, scalable, and sustainable in clinical practice[9]. IS also enhances evaluation frameworks, such as hybrid effectiveness–implementation study designs, that are well-suited to complex clinical environments.

## 3. IS Terminology

As IS continues to mature, so does the language used to describe its components. This section provides a foundational overview of key terms frequently encountered in IS literature and practice.

### Implementation Research (IR)

Implementation research refers to the scientific study of methods to promote the integration of evidence-based interventions into healthcare policy and practice settings to improve patient outcomes. It investigates why evidence-based innovations are not used, what barriers prevent their uptake, and what can be done to overcome these challenges. IR is used by some interchangeably with other concepts. Among them are the following two for which delicate differentiations have been proposed such as those seen below[33]:

- Implementation Science in which the aim is to produce generalizable knowledge
- Implementation Practice in which the main target is to produce local knowledge (see below)

### Medical Interventions/Innovations vs. Implementation Strategies

Medical interventions are diverse and many: they could be in form of either goods (e.g. pills) or care (AI-enhanced imaging services *delivered* to improve theranostics health outcome). In contrast, *implementation strategies* are the methods or techniques used to enhance the adoption, implementation, and sustainability of an intervention[39,40]. These might include training programs, clinical reminders, audit and feedback cycles, or co-design workshops with stakeholders. Strategies are selected or tailored based on their feasibility[41] and specific barriers identified through implementation research[42]. The contrast between medical interventions/innovations vs. implementation strategies has been graphically depicted in Figures 1 and 2.

When defining implementation science, some very non-scientific language can be helpful…

- The intervention/practice/innovation is **THE THING**
- *Effectiveness* research looks at whether **THE THING** works
- *Implementation* research looks at how best to help people/places **DO THE THING**
- Implementation strategies are the stuff we do to try to help people/places **DO THE THING**
- Main implementation outcomes are **HOW MUCH** and **HOW WELL** they **DO THE THING**

**Figure 1: Helpful summary of key language; From "Curran et al ( 2020). Implementation science made too simple: a teaching tool. *Implement Sci Commun*.;1(1):27."**

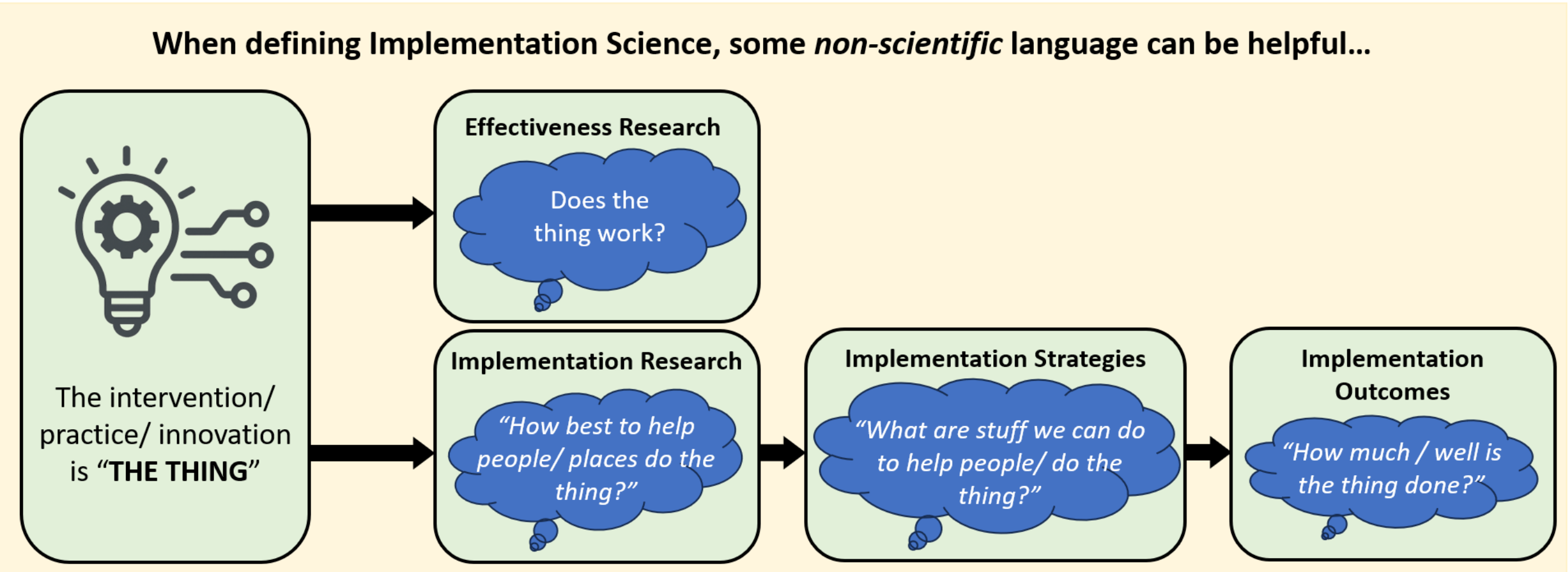

**Figure 2: An overview of core definitions in IS in non-scientific terms. "The thing" represents an intervention (e.g., an AI tool). The figure contrasts effectiveness research with implementation research, strategies, and outcomes**

### Implementation Outcomes vs. Clinical Outcomes

A crucial conceptual distinction in IS is the difference between **implementation outcomes** (what happens to the intervention in practice) and **clinical outcomes (what happens to the patient)**. While traditional clinical research focuses on issues such as improving treatment efficacy (e.g., tumor response, patient prognostication), IR assesses outcomes such as:

- **Adoption** (the initial uptake by individuals or departmental settings),
- **Scalability** (after the initial uptake, how it is scaled up and the uptake is followed by more number of people, departments, centers),

- **Fidelity** (the degree to which the intervention is delivered as intended),
- **Penetration** (integration within a service setting),
- **Sustainability** (the extent to which an intervention is maintained over time).

The primary objective in IS is to determine whether and how the intervention is being adopted and used in practice. While improved clinical outcomes are ultimately the goal, IS focuses on the success of adoption, delivery, and sustained use—all of which are essential precursors to realizing any clinical benefits that the tool may provide.

**Implementation Practice and Local Adaptation of Strategies**

Another frequently referenced concept is **implementation practice**—the real-world application of implementation principles in specific contexts, often requiring **local adaptation.** A single intervention may be effective in one setting but fails in another due to contextual differences in culture, infrastructure, or stakeholder engagement. This distinction reflects the difference between studying implementation in theory and applying it in practice, where success often depends on acquiring local knowledge and iteratively tailoring strategies to the certain clinical environment[43]**.**

**Theories, Models, and Frameworks (TMFs)**

These are foundational tools used in IS to plan, guide, and evaluate implementation efforts:

- **Theories** explain causal mechanisms or why things happen (e.g., Diffusion of Innovation Theory or others that similarly show what causes behavioral change among people)
- **Models** describe stages or processes of implementation or describe how implementation occurs step-by-step (e.g., Knowledge-to-Action or KTA Model)
- **Frameworks** organize what factors to consider and when and provide structures for describing or evaluating factors affecting implementation (e.g., the EPIS Framework: Exploration, Preparation, Implementation, Sustainment, versus CFIR—Consolidated Framework for Implementation Research).

Theories help articulate the relationships between key elements in an implementation model, while frameworks are often used to describe or categorize those elements systematically. Together, these tools offer structured language and logic that support both the design of implementation strategies and the evaluation of their success[44].

**4. IS vs. Knowledge Translation (KT)**

In the landscape of health research and policy, **KT and IS** are often mentioned together—and sometimes used interchangeably. While they share common goals, it is useful to distinguish their roles, particularly in the development and deployment of evidence-based tools such as AI applications in MI.

**Knowledge Translation** refers to the **dynamic and iterative process** of moving research into practice and policy[45].KT encompasses the synthesis, dissemination, exchange, and application of knowledge to improve health outcomes and optimize healthcare systems. In contrast, **IS** focuses more narrowly on the **scientific study of methods and strategies** that promote the systematic uptake of research innovations into the clinic. KT is often categorized into two broad approaches: End-of-Grant KT and Integrated Knowledge Translation (iKT)[46]**.**

**End-of-Grant KT**

This traditional model involves disseminating research findings after a project is completed, such as publishing papers, presenting at conferences, or sharing toolkits with stakeholders[47].While common in academic research, this model often falls short in complex, team-based environments like MIand theranostic applications, where integration of new AI-based solutions or imaging technology requires extensive cooperation, nuanced planning, clinical feedback, and iterative adjustment. Disseminating that occurs only at the end of a project may come too late to meaningfully shape the tool's relevance and usability.

**Integrated Knowledge Translation (iKT)**

In contrast, iKT emphasizes **early and ongoing engagement with knowledge users**—including clinicians, administrators, patients, and decision-makers—from the inception of the research process. Rather than positioning end users as passive recipients of innovation, iKT invites them to serve as co-designers and co-implementers from the start. This collaborative model fosters co-design, relevance, and readiness for adoption[48].

IKT is increasingly supported by research funders like the Canadian Institutes of Health Research (CIHR)[46], which emphasize the value of embedded research partnerships. In these models, knowledge users, such as clinicians, administrators or technologists, are not passive recipients of finalized solutions but active **co-creators** throughout the research process. Involving end users from the outset is critical; when solutions are developed in isolation and only introduced after years of work, stakeholders may question their relevance or practicality. Early and ongoing two-way engagement helps ensure that the research addresses real clinical needs and is positioned for successful implementation.

**Implications for AI and MI**

In the context of AI solutions for MI, applying KT and IS frameworks means engaging end-users such as physicians, technologists, and IT leaders not only during deployment, but keeping the in-the-loop from the early stages of algorithm design [49,50]. This includes defining clinically relevant needs, curating relevant data, and ensuring that outputs are interpretable and actionable within existing workflows. As an example of this, consider the evaluation of an AI-based algorithm to improve the quality of medical images. An evaluation that focuses on novelty may suggest that the use of metrics that quantify visual fidelity is sufficient. However, a stakeholder, such as the physician, may suggest that getting images that look good may not be sufficient, and what really matters to them is evaluation on the clinical task, and this may then fundamentally change the design of the AI algorithm to be optimized for clinical task performance.[51]

Further, another stakeholder may provide insights on how to use their domain knowledge to identify if the algorithm has underperformed.[52]Similarly, involving the stakeholders soon may help guide the validation of the AI algorithm such that the validation results provide more confidence to the physician. For example, a physician could help define the clinical task, process to collect patient data, procedure to define reference standard, and even how to quantify performance in a clinically meaningful and relevant way for validation of an AI algorithm.[53]

Engaging stakeholders early helps align research questions with real-world clinical priorities, facilitates context-sensitive implementation strategies, and improves the likelihood of sustainable integration. In some cases, individuals in operational roles may anticipate implementation needs or challenges more accurately than the research team itself. When involved from the outset, they can effectively help customize solutions and make real-world adoption feasible. By integrating KT and IS principles throughout the MI research pipeline, AI-based solutions in MI are more likely to achieve real-world adoption, close the evidence-to-practice gap, and contribute meaningfully to patient care and health system efficiency.

| Feature | Traditional KT (End-of-Grant) | Integrated KT (iKT) | Implementation Science |
| --- | --- | --- | --- |
| **When Stakeholders Are Engaged** | After results are finalized | From study inception through all phases | After or alongside effectiveness research |
| **Purpose** | Dissemination of findings | Co-creation of research questions and solutions | Study of methods to promote adoption |
| **Approach** | Linear (Research → Dissemination) | Iterative, Collaborative | Systematic, evidence-based |
| **Primary Output** | Papers, Reports, Toolkits | Solutions aligned with real-world needs | Strategies, frameworks, implementation outcomes |
| **Primary Focus** | Communication of knowledge | Integration of knowledge into practice | Evaluation of adoption, fidelity, sustainability |
| **Key Users** | Researchers, Policy Brief Authors | Clinicians, Administrators, Patients (as co-designers) | Implementation teams, System stakeholders |
| **Example in Imaging Context** | AI model developed, published, and shared | AI model developed with clinician input from start | Evaluating best way to deploy AI into imaging workflow with clinician input from start |

**Table 1:** ***Comparison of traditional end-of-grant knowledge translation, integrated knowledge translation (iKT), and IS. iKT bridges the gap between evidence generation and implementation by involving knowledge users throughout the research process. This integrated approach is especially valuable in complex domains such as AI-driven MI, where contextual understanding and adaptability are essential.***

## 5. AI Evaluation vs. IS in MI

There is a key distinction between **AI evaluation** and **IS (Figure 3)**.

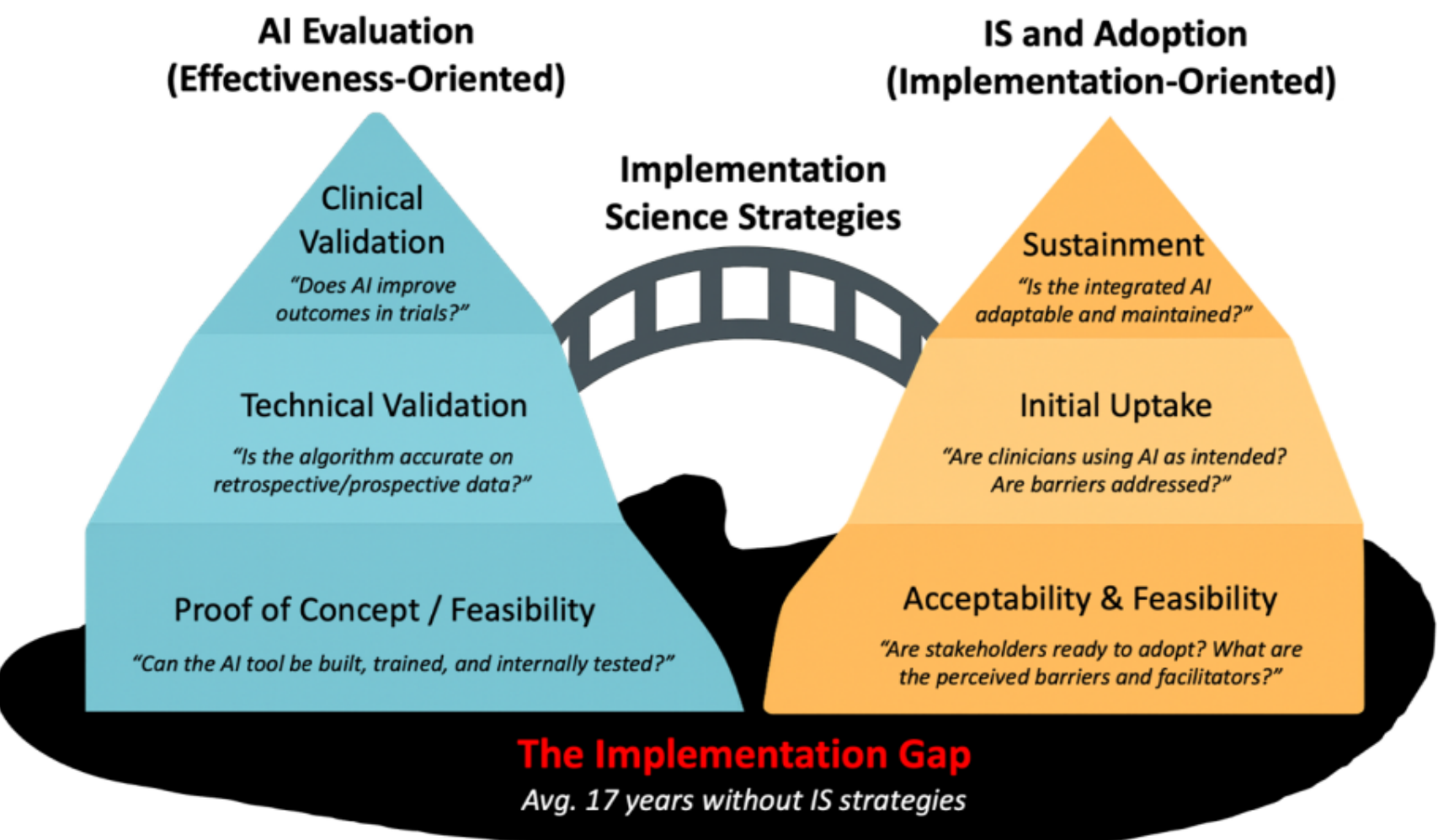

**Figure 3: "Bridging the Gap: From AI Evaluation to Real-World Implementation in MI": A layered model illustrating the distinction between AI evaluation and real-world implementation stages. While evaluation assesses technical and clinical**

The former typically focuses on **developing, testing, and validating algorithms** using retrospective or prospective datasets. As suggested in the RELAINCE guidelines,[54] this evaluation can have multiple classes. Performance evaluation of AI-based methods in MI typically involves visual fidelity metrics (e.g. RMSE or SSIM) or technical measures (e.g. Dice Similarity Coefficient). The goal of such evaluation is to provide proof-of-concept validation of the method and illustrate the technical innovation. Next, to determine performance on clinical task, objective task-based evaluations are needed[53,55]. Such evaluations determine how well AI tools perform on specific clinically relevant tasks such as those of detection and quantification. For example, performance on detection tasks can be measured using metrics such as the Receiver Operating Characteristic (ROC) analysis obtained from model observer and human observer studies. More advanced evaluation quantifies whether the AI algorithm can improve performance in making clinical decisions, including diagnostic, prognostic, predictive, and therapeutic decisions for primary endpoints such as improved accuracy or precision in measuring clinical outcome.

Yet even when AI tools are deployed, a substantial gap can remain between technical success and clinical adoption. Demonstrating that an algorithm works in principle is only part of the challenge—real-world adoption depends on whether the tool fits into various stages of MI workflows, addresses clinician needs, and overcomes practical barriers. Imaging workflow in radiology has been indicated to include seven stages and AI tools have been developed for improvement in entire workflow stages[56].

This is where IS becomes essential. It does not necessarily seek to demonstrate whether AI works, but investigates why it is—or is not—being used, and identifies the strategies necessary to support successful integration into clinical care. IS shifts the focus from "Does it work?" only to also "How do we make it work in practice?"

In many cases, the lack of adoption stems from organizational, behavioral, or system-level barriers—issues rarely addressed in traditional AI research. For instance:

- Bureaucratic and other difficulties in obtaining regulatory approval for AI tools that could be integrated in multiple centers into varying software packages used in their differing radiology departments in multi-centers
- AI tools may require **workflow redesign** that disrupts existing roles
- Clinicians may lack **trust** in or understanding of the algorithms
- Institutions may not have the **infrastructure** or **IT support** to deploy models
- There may be **no clear incentives** for use, or misalignment with reimbursement structures.

These barriers are particularly salient and costly in MI[57], a domain already characterized by complex scheduling, diagnostic agent logistics, high equipment expenses, and interdisciplinary decision-making. AI must integrate seamlessly into this environment. It must not simply be technically sound, but also **acceptable, feasible, and sustainable** in daily practice.

To bridge this gap, IS offers tools such as:

- **Formative assessments** to understand barriers and facilitators to AI use
- **Implementation frameworks** (e.g., CFIR, EPIS) to guide deployment
- **Implementation strategies** such as training, workflow integration, and stakeholder engagement
- **Hybrid research designs** (discussed in Section 6) to simultaneously assess effectiveness and implementation outcomes.

In summary, while AI evaluation tells us whether a tool *can* work, IS tells us how to ensure it *does* work—consistently, equitably, and at scale.

**6. Hybrid Effectiveness-Implementation Designs**

Traditional clinical research has often followed a linear pathway: establish efficacy in controlled settings, then effectiveness in real-world scenarios, and finally consider implementation. While logical in theory, this sequence is often too slow and poorly aligned with the pace of innovation in domains such as AI in MI. To address this disconnect, implementation scientists have developed hybrid effectiveness-implementation designs—a methodology that allows researchers to evaluate both clinical outcomes and implementation processes concurrently[58,59].

In the context of AI, this approach is especially important. The traditional separation between effectiveness research and implementation planning is no longer feasible. AI tools may evolve rapidly, and waiting until after clinical validation to consider integration will result in missed opportunities, outdated solutions and low adoption.

Hybrid designs help answer two types of questions **simultaneously**:

- Does the intervention (e.g., an AI-based MI interpretation tools) improve outcomes?
- How well is the intervention being adopted, delivered, and sustained in practice?

**Three Types of Hybrid Designs**

Hybrid studies are categorized into **three types**, depending on the primary and secondary aims. In Type 1 designs, researchers focus primarily on whether an AI tool works as intended but begin to gather data on contextual factors that may influence adoption. Type 3 designs, by contrast, are most appropriate when the clinical benefits of the tool are already established, and the goal is to evaluate which implementation strategies are most effective in promoting its use. In the middle are Type 2 designs in which equal emphasis is placed on both sides.

Hybrid designs are particularly well-suited to **AI-enabled MI**, where:

- Algorithms are often developed alongside users,
- Clinical and technical success depend heavily on integration into real-world workflows,
- Evidence must be both **robust** and **relevant** to multiple settings.

**Why Hybrid Designs Matter for AI in MI**

AI tools in MI rarely follow a clean, linear path from development to clinical adoption. Instead, their journey typically involves iterative refinement, local adaptation, and continuous feedback—making it essential to assess both their impact and feasibility in tandem.

Hybrid designs help close the gap between AI development and adoption by enabling research teams to:

- Optimize AI models and their delivery strategies in parallel,
- Identify barriers to adoption in real time,
- Generate real-world evidence to inform scale-up across diverse clinical settings.

These designs also align closely with iKT principles (as discussed in Section 4), empowering researchers and clinicians to co-develop solutions that are both technically sound and practically implementable from the outset.

| Hybrid Design Type | Primary Aim | Secondary Aim | Study Scenario | Example in IMAGING /AI Context |
|---|---|---|---|---|
| **Type 1** | Evaluate **clinical effectiveness** | Explore **implementation context** (e.g., barriers, readiness) | You have a promising intervention but limited data on real-world use | Test if an AI tool improves IMAGING diagnostic accuracy while noting clinician trust, workflow fit, and infrastructure constraints |
| **Type 2** | Evaluate both **effectiveness and implementation** equally | Dual focus on impact *and* real-world delivery | You want to assess both impact and uptake during early stages of rollout | Simultaneously test AI's performance and the feasibility of adoption across multiple IMAGING centers |
| **Type 3** | Evaluate **implementation strategy** | Observe clinical effectiveness of an already validated tool | Intervention is already effective; focus is on best ways to scale and sustain use | Compare two strategies (e.g., centralized training vs. embedded champions) to support adoption of an IMAGING-AI diagnostic tool across hospital networks |

**Table 2:** ***Overview of hybrid effectiveness-implementation design types: These designs allow simultaneous investigation of clinical impact and real-world feasibility or strategy effectiveness. Choice of type depends on the maturity of the intervention and the primary focus of the study. In AI-enabled MI, hybrid designs offer a structured approach to evaluate both technical outcomes and adoption success.***

## 7. Barriers to AI Implementation

A set of **multi-level barriers** prevent AI tools from being used at scale in real-world environments. A comprehensive framework outlined by the National Academy of Medicine[9]identifies barriers across four domains and eight dimensions, as demonstrated in Figure 4.

**AI Implementation Challenges: 4 Domains**

**1. Reason to Use**
- Misalignment between AI tools and clinical priorities
- Lack of clear incentives for adoption

**2. Means to Use**
- Inadequate infrastructure
- Hard/software challenges
- Data integration challenges
- Insufficient IT support

**3. Methods to Use**
- Poor interpretability of AI outputs
- Workflow incompatibility
- Lack of explainability

AI output

**4. Desire to Use**
- Lack of trust in AI
- Fear of workflow disruptions
- Medicolegal implications

**Figure 4: These domains range from infrastructure limitations and workflow incompatibilities to deeper cultural and psychological barriers, such as mistrust or *resistance* to change. What becomes clear from this framework is that technical merit alone is not enough. AI solutions must not only demonstrate efficacy but also resonate with clinical goals, fit seamlessly into workflows, and earn the trust and acceptance of end users.**

**Integrating HCI Frameworks in MI: Towards Usable AI**

AI holds great promise for improving MI workflows. Yet, in practice, the integration of AI tools into clinical practice remains uneven. One major reason is the lack of attention to human and organizational factors during development and deployment. HCI, and in particular User-Centered Design (UCD), offers a robust yet often underutilized framework for addressing these challenges[60].

In a nutshell, HCI focuses on designing computing systems that align with how humans think and work, ultimately improving user experience and practical outcomes. In healthcare, especially in MI, HCI facilitates smoother integration of AI tools into clinical workflows by aligning system design with users' mental models and preserving human interests. This not only improves usability but also builds trust among diverse stakeholders. Although HCI is an interdisciplinary field (Figure 5)[61] with strong roots in design and cognitive sciences, its key framework -User-Centered Design (UCD), remains underutilized in health contexts, particularly where complex, multi-user environments like MI are involved.

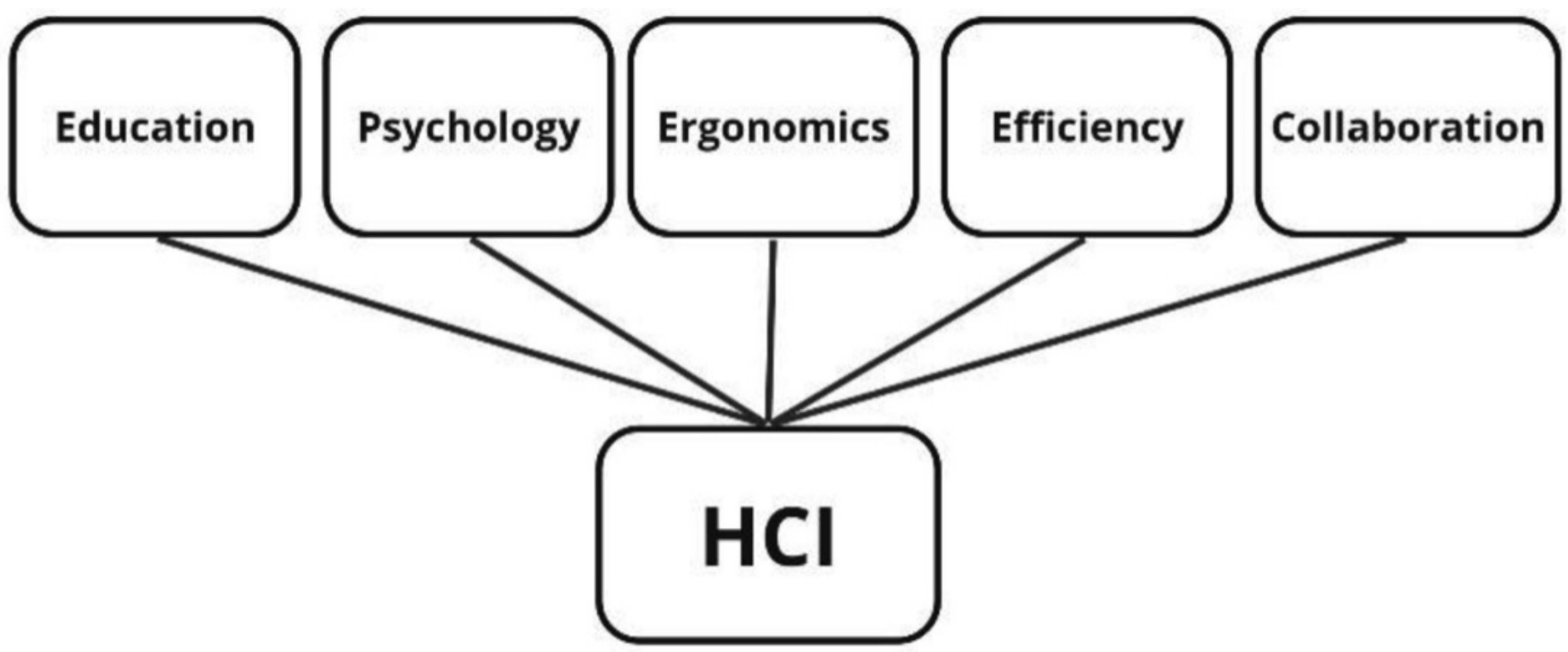

Figure 5: Different contributing components for HCI (from: Meher Langote, Saniya Saratkar, Praveen Kumar, Prateek Verma, Chetan Puri, Swapnil Gundewar, Palash Gourshettiwar. Human–computer interaction in healthcare: Comprehensive review[J]. AIMS Bioengineering, 2024, 11(3): 343-390. doi: 10.3934/bioeng.2024018)

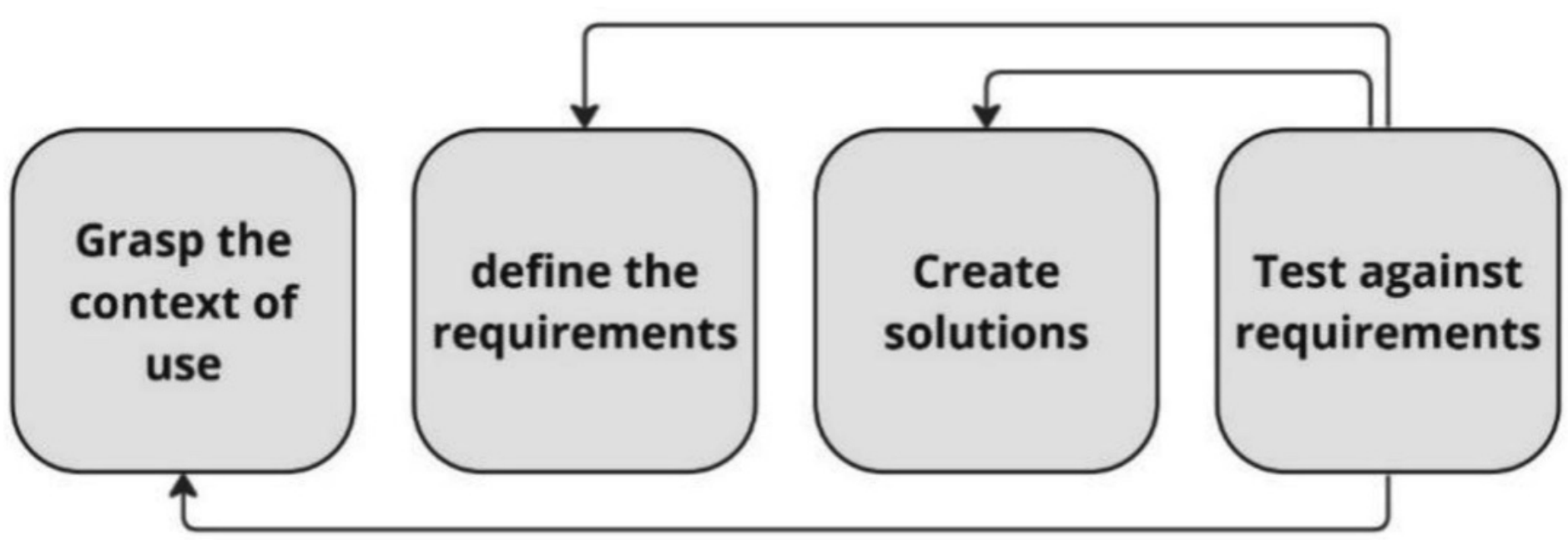

Figure 6: Different phases of user-centric design (from: Meher Langote, Saniya Saratkar, Praveen Kumar, Prateek Verma, Chetan Puri, Swapnil Gundewar, Palash Gourshettiwar. Human–computer interaction in healthcare: Comprehensive review[J]. AIMS Bioengineering, 2024, 11(3): 343-390. doi: 10.3934/bioeng.2024018)

Widely adopted in software engineering, UCD emphasizes iterative development (Figure 6)[61], real-time feedback, and responsiveness to the end-user's context. In MI, this means designing tools not just for clinicians, but **with** themto ensure usability, interpretability, and workflow fit from the outset.Despite growing recognition of these needs, many AI adoption efforts in MI often remain concentrated at the **perceptual stage**—exploring clinicians' general attitudes toward AI rather than investing in sustained engagement before, during, and after deployment. This limited scope, frequently motivated by the need to justify funding or organizational support, results in tools that are theoretically sound but poorly adapted to clinical realities[62]. In MI where workflows are tightly choreographed and diagnostic accuracy is paramount, the cost of this misalignment is particularly high.

**Barriers Rooted in Culture, Context, and Collaboration**

This disconnect is further deepened by two persistent challenges(Figure 7). The first is **territoriality between disciplines**: despite calls for more integrated collaboration, the divide between computer science and medical practice remains marked. Specialists in these domains often approach problems with fundamentally different assumptions, methods, and goals, which makes interdisciplinary synergy difficult. Such divides have proven in many investigation to be remarkably durable, including in the studies elsewhere by us [22,29].

The second is a **lack of clinical contextualization** in the development of AI tools. Incentives in both academia and industry tend to favour positive outcomes and model-centric contributions, which limits attention to usability, negative results, or failed implementations[63,64]**.** In MI—where multimodal data interpretation, patient throughput, and radiotracer logistics add layers of complexity—tools that ignore this context are likely to fall short of their intended impact.

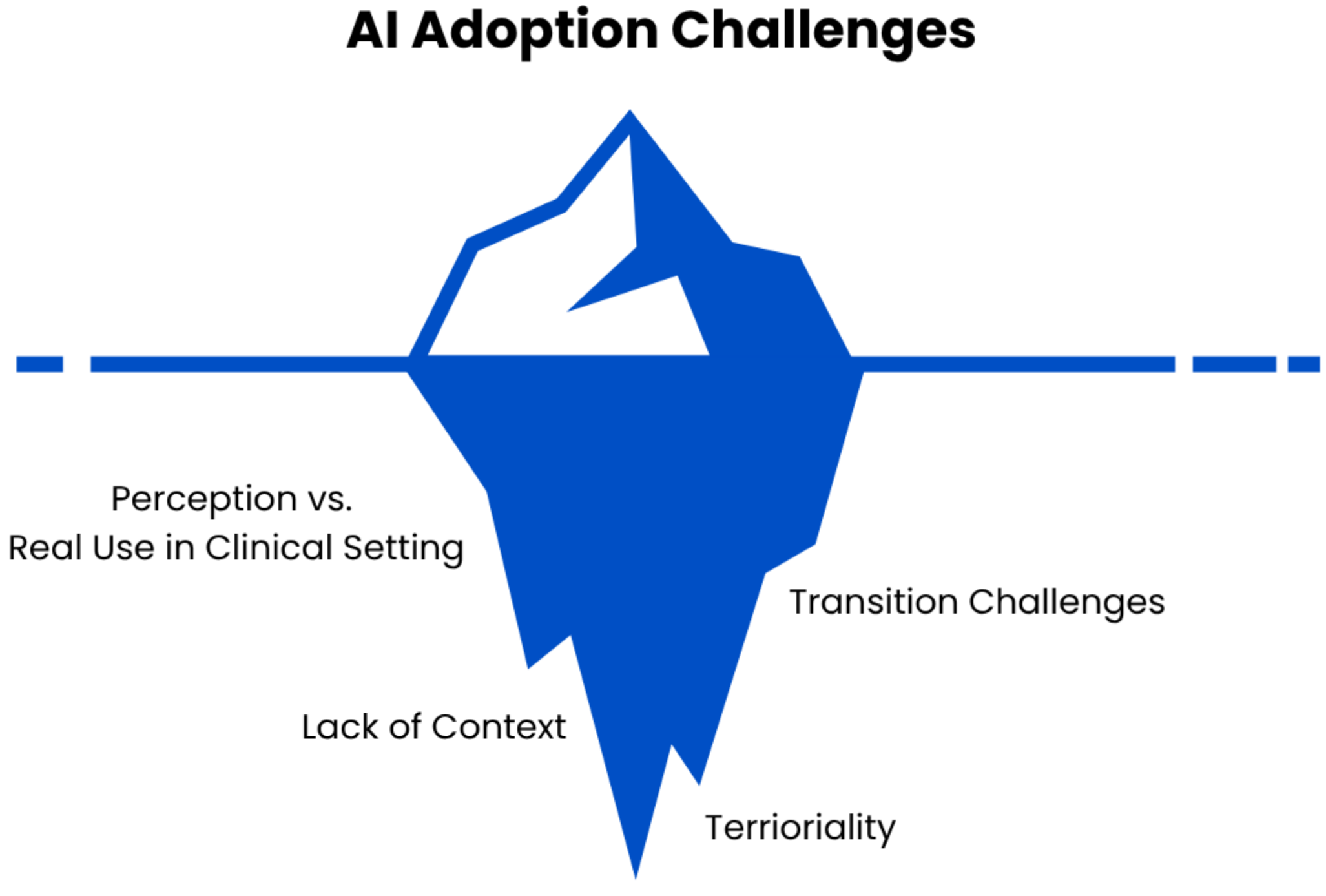

**Figure 7: Challenges to adoption of AI. In bold are two major issues. HCI/UCD solutions hold significant potentials in tackling key barriers.**

These issues come into sharp relief during the transition from controlled research environments to everyday clinical use. AI models that perform well on retrospective datasets **frequently stumble in operational settings**, where they must integrate with legacy systems, align with radiologists' mental models, and support real-time decision-making. In computer-based technology adoption, where decisions often carry high diagnostic and therapeutic stakes, **the perceived lack of transparency in facilitators of decision-making can erode trust**[9]. Coupled with minimal usability testing and insufficient workflow adaptation, such systems are prone to abandonment[1,10,17,21,29].Clinicians, already managing high cognitive and emotional workloads, may resist adopting tools that introduce uncertainty or require significant behavioural change[2229].

To address these issues, a shift is needed—from designing *for* clinicians to designing *with* them. This is where HCI, and particularly UCD, offers crucial guidance. UCD frameworks advocate for continuous user involvement at every stage: from identifying the problem space to co-developing prototypes, and from refining based on usability testing to monitoring post-deployment use. In the context of MI, this means involving physicians, radiologists, technologists, and even patients in shaping how AI systems are designed, validated, and deployed. The goal is not merely to increase user satisfaction, but to create AI tools that are seamlessly embedded in the clinical ecosystem—responsive to its constraints, aligned with its values, and trusted by its practitioners.

Overall, while MI exemplifies the technical promise of AI, it also illustrates the sociotechnical barriers that continue to impede real-world adoption. Overcoming these challenges will require more than algorithmic refinement—it will demand a broader reorientation toward co-design and human-centered thinking. By applying principles from HCI and UCD, we can move closer to an AI future in MI that is not only innovative, but also meaningful, usable, and lasting.

**Application to MI**

In MI, the implementation challenge is magnified by several unique contextual factors:

- **Workflow complexity**: PET involves multiple handoffs (e.g., image acquisition, reconstruction, interpretation, reporting), each of which could be disrupted—or enhanced—by AI.
- **Specialized expertise**: Clinicians and technologists may be unfamiliar with AI's capabilities or uncomfortable interpreting its outputs.
- **High stakes**: MI is often used in oncology and neurological diagnostics, where diagnostic uncertainty carries significant consequences.

There are **seven potential entry points for AI in the MI/radiology workflow**[56]:

1. **Imaging order generation**
2. **Patient scheduling**
3. **Image protocoling**
4. **Image acquisition and reconstruction**
5. **Image interpretation**

6. **Report generation**
7. **Report communication**

Each entry point represents both an **opportunity for AI augmentation** and a **site for potential barriers**—from resistance by technologists and radiologists to lack of integration with PACS/RIS systems or administrative bottlenecks.

Even in domains where the clinical value of AI is widely recognized, such as exam scheduling, significant barriers persist. The challenge often lies not in technical feasibility, but in institutional readiness and cultural alignment.

**Toward Barrier-Responsive Implementation**

Understanding barriers is not merely an academic exercise; it forms the basis for designing effective **implementation strategies**. For example:

- Low trust → **Training, co-design workshops, and transparent AI explanations**
- Workflow mismatch → **Redesign of clinical pathways with frontline stakeholder input**
- Infrastructure gaps → **Investment in interoperable platforms and IT support**

By systematically identifying and addressing these barriers, IS helps convert *AI potential* into *clinical reality*. In the MI context, this is essential—not only for improving diagnostic care, but also for ensuring that innovations benefit patients across institutions and populations.

**8. Collaborations and Partnership**

Successful implementation of AI in MI depends not only on evidence and innovation but on **people, relationships, and shared ownership**. IS acknowledges that achieving real-world impact requires **multisectoral, interdisciplinary collaboration** among a broad spectrum of stakeholders. One of the key insights from IS is that no single group—whether researchers, developers, or clinicians—can drive adoption alone. Effective implementation requires strong partnerships between knowledge creators, users, and brokers (see Table 3 and Figure 8), built on shared goals and collaborative engagement.

| Role | Description |
| --- | --- |
| **Knowledge Creators** | Researchers, engineers, data scientists who develop AI tools or generate evidence |
| **Knowledge Users** | Clinicians, administrators, health professionals who apply tools in practice |
| **Knowledge Brokers** | Individuals or teams who bridge the gap, facilitating communication, aligning incentives, and co-designing solutions |

**Table 3: Three Key Stakeholder Roles in Implementation**

While the concept of knowledge brokers is sometimes unfamiliar in clinical settings, their function is critical. Much like intermediaries in finance or real estate, brokers help translate needs between groups, surface latent barriers, and align priorities before formal implementation even begins. They are often the connective tissue that ensures innovations are not only built but adopted meaningfully.

Importantly, these roles are not fixed. One individual or team may play multiple roles over the course of a project. For example, a physicist might evolve into a knowledge broker by facilitating communication between engineers and clinicians. Similarly, a clinician may become a knowledge champion by advocating for AI integration in departmental planning.

### Co-Creation and Mutual Transformation

Modern implementation frameworks strongly emphasize co-creation—a collaborative model in which knowledge users and creators jointly design solutions, share insights, and challenge each other's assumptions. This approach not only enhances the relevance and sustainability of innovations but often leads to personal and professional transformation[65].

Beyond just a technical process, co-creation can also lead to personal and professional transformation. Engaging in co-creation often reveals blind spots in one's assumptions, whether from the perspective of the researcher or the clinician. This is especially critical in AI implementation, where clinicians must be involved not only in validating algorithm performance, but also in shaping how and where they are deployed. The best implementation strategies often emerge not from research teams alone, but from shared decision-making among all parties affected by the change.

### The Risk of Isolated Efforts

Without collaboration, even the most elegant solutions may fail to reach patients if developed in isolation. This phenomenon is often depicted in implementation frameworks that illustrate the cumulative drop-offs in engagement at each level (e.g., providers not trained, trained providers not delivering, delivered interventions not received). Various iKT guidelines highlight that even small breakdowns at different stages of delivery can result in only **a fraction of intended benefit** reaching its targets[46]**.**

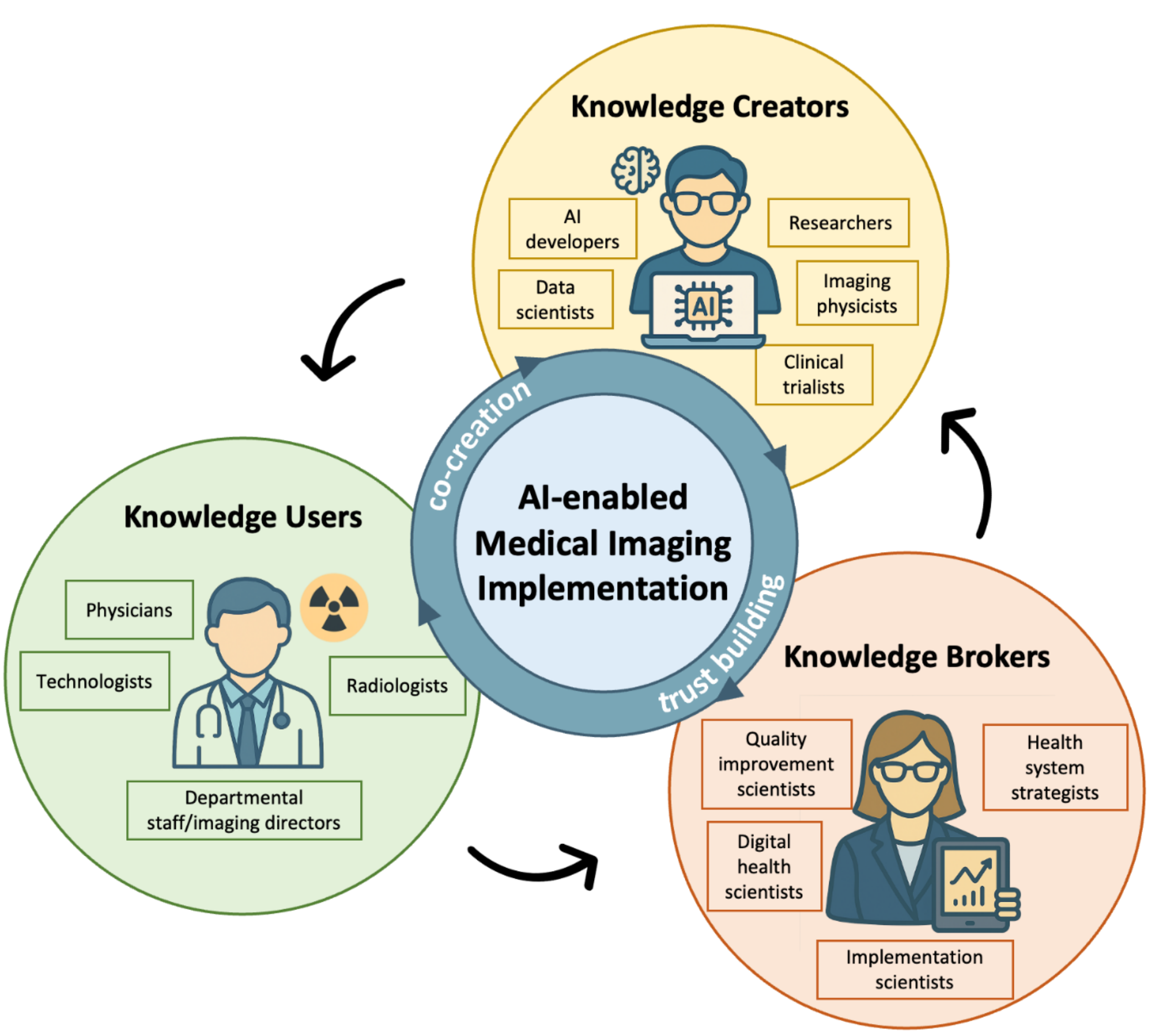

Figure 8: A partnership ecosystem map illustrating the interconnected roles of knowledge creators, users, and brokers in implementing AI solutions in MI. Effective implementation requires mutual engagement, feedback loops, and shared responsibility

IS is not simply about strategy—it is about shared responsibility. As AI continues to transform the field of MI, researchers, clinicians, administrators, and system designers must move forward together. The future of innovation lies not just in technical breakthroughs, but in building trust, sustaining partnerships, and co-creating change.

## 9. Conclusion: Bridging Innovation and Impact in MI

The integration of AI solutions into MI holds transformative potential. However, realizing this promise demands a paradigm shift in how we evaluate, adopt, and sustain innovation in clinical settings. While the development and technical validation of AI tools continue to advance rapidly, their real-world clinical utility remains constrained by underdeveloped implementation pathways. IS offers the blueprint to bridge this gap.

In this work, we discussed various key IS terms, models and frameworks, and how IS connects with knowledge translation paradigms, and can be combined with human-computation interaction (HCI) principles. We also emphasized the value of employing hybrid effectiveness-implementation designs, understanding and navigating implementation barriers, and leveraging iKT frameworks. These approaches recognize that adoption of innovation is not linear; rather, it occurs through complex, context-specific, and stakeholder-driven processes. The MI AI implementation journey must be collaborative—guided by partnerships among developers and clinicians, health system leaders, and patients. As MI becomes increasingly data-intensive and AI-enhanced, success will depend on how well we align our adoption strategies with the principles of IS. Only then can we ensure that AI tools are not just technically impressive, but equitably and effectively adopted, sustainably used, and truly impactful in improving patient care.

## Acknowledgements

This work was supported by the Tehran University of Medical Sciences and by the Natural Sciences and Engineering Research Council of Canada (NSERC) Discovery Horizons Grant DH-2025-00119.